%Paper: astro-ph/9506010
%From: aikman@dao.nrc.ca (Chris Aikman)
%Date: Fri, 2 Jun 95 00:29:51 PDT

\magnification=\magstep1
\baselineskip 19pt
\centerline{\bf QSO HOST GALAXIES AT Z=2.3}
\vskip 20pt
\centerline{J.B.Hutchings\footnote{$^1$}{Guest observer, Canada France Hawaii
Telescope, which is operated by NRC of Canada, CNRS of France, and the
University of Hawaii}}
\centerline{Dominion Astrophysical Observatory, National Research Council
of Canada}
\centerline{5071 W. Saanich Rd, Victoria, B.C., V8X 4M6, Canada}
\vskip 20pt
\centerline {\bf Abstract}
\vskip 5pt
Images are discussed of 6 QSOs at z=2.3, one QSO-like IRAS source at z=2.3,
and one QSO at z=1.1, taken with resolution 0.6 to 0.9 arcsec. 5 of the
QSOs are radio-quiet. All QSOs except one are just resolved, while the
IRAS source has definite structure. In some cases, part of the QSO fuzz
appears to be a close companion rather than a concentric host galaxy.
The luminosities implied for the hosts or companions are typical of
bright galaxies with young hot star populations. Radio-quiet QSOs appear to
have host galaxies less luminous by $\sim$2 magnitudes, than radio-loud QSOs.

\vfill\eject
\centerline{\bf 1. Introduction}
\vskip 10pt
Host galaxies of QSOs of all types have been resolved at redshifts up to
$\sim$0.5 (e.g. Hutchings and Neff 1992). At redshifts 0.5 to 0.9, host
galaxies of radio-loud QSOs are resolvable with good signal and image
quality ($<$0.7 arcsec: see Hutchings 1992). At high redshifts ($\ge$2),
only those of very high radio luminosity have been resolved
(Lehnert et al 1992). Lowenthal et al (1995) report marginal K-band detection
of radio-quiet host galaxies at z$\sim$1 but no detection at z=2.5.
In addition to the host galaxies themselves,
radio-loud QSOs often have extended emission-line gas, sometimes of large
size and luminosity, and often associated with the extended radio structure
(Heckman et al 1991). It thus appears that at higher redshifts, radio-loud
QSOs are hosted by exceptionally luminous galaxies, while the majority
of (radio-quiet or lower radio-luminosity) QSO host galaxies have been too
faint to detect. The results reported in this paper resolve for the first
time, the host galaxies of high redshift radio-quiet QSOs.

    The observations are already reported by Hutchings, Crampton and
Johnson (1995) and Hutchings (1995), in discussions of the environments
of these objects. The data were obtained using the High Resolution
Camera of the CFHT. This covers a field of $\sim$2 arcmin with 0.11 arcsec
pixels, and rapid tip-tilt guiding. Exposures were 600 sec in R-band for
all objects and also in I band for the z=1.1 object. These correspond to
rest wavelengths of 2100A at z=2.3 and 3300A and 3800A at z=1.1. The image
FWHM was in the range 0.6 to 0.9 arcsec, which is significantly worse than
the 0.4 arcsec obtainable in the best conditions. The image width at 1\%
of peak (see below) is $\sim$3 arcsec. The detector was a Loral CCD with
$\sim$6e read noise, and gain 1.9 e/ADC unit.
\vskip 10pt
\centerline{\bf 2. Data reductions}
\vskip 10pt
   The observing conditions were photometric with variable seeing, and
variable sky brightness due to moonlight. Photometric
standard fields were observed on each of the two nights of observation, in
April 1993. In each standard field about 10 stars were used to generate
the calibration curve: resulting magnitudes for the QSOs are estimated to
be accurate to $\pm$0.1 mag.
The point spread function (PSF) differed significantly between
exposures, but not across any individual image.
Rapid guiding was done on a bright star within $\sim$1 arcmin
in all fields. In each image there was one or more bright star and several
faint stars (comparable with the QSO) to use to generate PSFs for each image.
PSF stars were chosen to be free of close galaxy companions, cosmic rays, and
detector flaws. In a few instances, cosmic ray signals of a few pixels near
to the PSF stars or the QSO were edited out and replaced by values close
to the local mean. The processed image sky signal is flat to well below
1\% over the small fields of interest (20 arcsec or less).

   The QSO images were not saturated and were measured using IRAF aperture
photometry. Some of the bright PSF stars were saturated in their centres
and these pixels were not used. The Images of the QSOs and the PSF stars were
fit using the `ellipse' task of STSDAS, which also estimates RMS errors for
each
value. The PSF peaks were derived from the fainter stars and the outer parts
using the bright stars. In all cases there was excellent agreement between
PSF profiles in the same image. Figure 1 shows some PSFs and the scatter
among the individual stars used.

    Figure 2 shows the QSO profiles fitted with PSF profiles which are matched
at the peak signal. The principal uncertainty in the process is the sky
level to subtract, since we are measuring the profiles down to signal
levels within one or two counts of the sky, which had a typical value of 150.
The sky photon noise is the principal noise source. The sky level was measured
by image statistics near the stars, and by the `rimexam' task in IRAF; in all
cases values were tried over a range of 5 about the best value, to check the
sensitivity of the results. In most cases, the correct sky value was clearly
indicated by the profile signal stabilizing around zero far from the object,
where there are no other objects seen in the image. In no case was the
apparent resolution of the QSO dependent on the sky level chosen within this
range.

    In all cases except one, the QSO profile is broader than the PSF at
radii above the 1 -2 arcsec range. Beyond about 5 arcsec, the signal level
is too low to measure, except where there are obvious faint galaxies near
the QSO. While a number of individual profile points lie within their error
bars
of the PSF profile, there are many that lie systematically 1 - 2 $\sigma$
above the PSF, and none that lie more than 1 $\sigma$ below. Thus,
the cumulative result is that
the QSO images are generally extended with faint signal.

    Considering the faintness of the signal and the short range of radius
over which it is detected, no attempt was made at measuring the shape of
the PSF-subtracted
profile. The integrated extended light was measured and is given in Table 1.
The 1$\sigma$ uncertainty in the ratio of nuclear to resolved flux
is a factor 1.5, or about 0.5 in host galaxy magnitude, at host galaxy
magnitudes $<$22. For the fainter ones the uncertainty rises to $\sim$twice
this factor.

\vskip 10pt
\centerline{\bf 3. Individual objects}
\vskip 10pt
\bf 0820+296. \rm This radio-loud QSO has several companions nearby (see
Hutchings 1995) and they show up in the profile beyond 4 arcsec. The
extended light within this radius appears to be azimuthally spread.

\bf 10214+4724. \rm This IRAS source with QSO-luminosity has been resolved
before (e.g. Soifer et al 1992). The object has an elliptical inner
shape and a curved tail emerging from the long side of the ellipse.
Figure 3 shows contours from the CFHT image.

   The central resolved elliptical host has half the luminosity of the
unresolved nucleus, and the arc has comparable luminosity to the host.
The arc extends more than 4 arcsec in this image (20 Kpc at high redshift).
Soifer et al suggest that the arc is a separate and possibly disconnected
galaxy as its peak flux lies well away from the central source. The image
in Figure 3 shows that the arc does extend to within 1 arcsec of the
nucleus, and thus seems very likely to be connected with the source -
probably an tidally extended companion in the process of merging.

   Broadhurst and Lehar (1995) have modelled this object as a seyfert
galaxy lensed by an elliptical galaxy at z$\sim$1. In the present observations
only the central `nuclear' object is seen in B and NB (rest Ly$\alpha$) images
taken at the same time. This may further test the lensing model.

\bf 1232+134. \rm This QSO has faint resolved fuzz visible below 23.8
mag/arcsec$^2$. Some guiding jitter affected the central pixels, equally
in QSO and PSF.

\bf 1246+005. \rm This is the only QSO that shows no resolved luminosity.
The 1$\sigma$ upper limit for undetected luminosity is 1/50 of the QSO
flux, as given in Table 1.

\bf 1338+277. \rm This is the most clearly resolved QSO image, and appears
fuzzy on the screen. The QSO is also the faintest of the sample. There is
a faint compact knot as well, that lies 1.2 arcsec N of the nucleus.

\bf 1632+391. \rm This is the only QSO at lower redshift, and is radio-loud.
It has centrally distributed resolved light. There is a large nearby
galaxy (see Hutchings, Crampton and Johnson 1995) that was edited out of
the image or azimuthally ignored in the measurements.

\bf 1641+395. \rm The resolved luminosity lies principally to one side
and probably includes a very close companion galaxy. The QSO lies in a
very compact group of faint galaxies (Hutchings 1995).

\bf 1641+410. \rm This QSO also lies in a compact group of companions
(Hutchings 1995).
Some ($\sim$60\%) of the resolved light may come from a non-centred
faint companion.
\vskip 10pt
\centerline{\bf 4. Discussion}
\vskip 10pt
  The results indicate that we have just resolved the host galaxies of
the sample of high redshift QSOs observed with the HR Camera.
Table 1 shows the measured quantities from the PSF subtraction. Since the
limiting isophote level and the seeing and guiding are different for each
image, the results probably depend on these quantities. A plot of limiting
isophote against nuclear to resolved flux suggests that higher ratios
can be detected with fainter isophotes, as expected. We might also
expect the resolved host flux to depend on the nuclear fux, but this
does not show up in a plot. Thus, we probably see a real spread in host
galaxy luminosities.

   It is interesting to note that the host galaxies of the radio-loud
QSO 0820+296 and the IR source are the brightest. The other radio-loud
QSO is at z = 1.1, and at z = 2.3 would be at the limit of resolution
with these data. 10214+4724 is also a radio source: all 3 radio-loud objects
in the sample have fluxes of 0.4 to 0.5 Jy at 5 GHz, which corresponds
to the fairly high luminosities of (log) 26.5 - 27.5 W/Hz. However, they
are an order of magnitude less luminous than typical high redshift 3C
sources.

   In Table 1 the absolute magnitudes are given for the unresolved and
resolved flux, for H$_0$=100, q$_0$=0.5. No k-correction is applied to
any of these values. The rest wavelength of the observations at the QSO
is 2100A (3300A for 1632+391), so the k-correction depends greatly on
the stellar population. If the galaxies have present-epoch populations,
the k-corrections must be 6 or more magnitudes, making the host galaxies
extraordinarily luminous. If they consist of type B0 stars, they are
2 magnitudes \it less \rm luminous, and if they consist of A0 stars,
they are 0.5 magnitudes more luminous. Thus, the detected host galaxies
have the luminosity of normal bright galaxies if they have a young population
of stars. This seems a plausible scenario, and implies that the QSOs
live in new galaxies or extreme starbursting galaxies at this redshift.
This is consistent with the results on the companion galaxies discussed
by Hutchings et al (1995).

   Lehnert, Heckman, Chambers, and Miley (1992) found resolved light
about 2 magnitudes brighter than these radio-quiet QSOs, in a sample of
radio-loud QSOs. Hutchings, Ellingson, and Ozard (1992) also resolved
a radio-loud QSO at z=2 with a similarly luminous host galaxy.
Two of the 3 radio-loud objects in this sample, have
host galaxy magnitudes comparable with theirs. Thus, it appears that
radio-quiet host galaxies are considerably less luminous than radio-loud
ones - at this rest wavelength. This result is consistent with the lack of
detection of radio-quiet host galaxies with lower optical resolution
than the present sample. Clearly, better results and colour measurements
should be possible with deep exposures in excellent seeing conditions.
\vfill\eject
\centerline{\bf References}
Broadhurst T., Lehar J., 1995 (preprint)

Heckman T.M., Lehnert M.D., van Breugel W., Miley G.K., 1991, ApJ, 370, 78

Hutchings J.B., Ellingson E., Ozard S., 1992, PASP, 104, 1230

Hutchings J.B., 1992, AJ, 104, 1311

Hutchings J.B., Neff S.G., 1992, AJ, 104, 1

Hutchings J.B., Crampton D., Johnson A., 1995, AJ, 109, 73

Hutchings J.B. 1995, AJ, in press

Lehnert M.D., Heckman T.M., Chambers K.C., Miley G.K., 1992, ApJ, 393, 68

Lowenthal J.D., Heckman T.M., Lehnert M.D., Elias J.H. 1995, ApJ, 439, 588

Soifer B.T., Neugebauer G., Matthews K., Lawrence C., Mazzarella J., 1992,
ApJ, 399, L55
\vskip 15pt
\centerline{\bf Captions}
\vskip 10pt
1. Examples of comparison of PSFs for stars in a single image. Different
symbols
distinguish the stars. The dashed line indicates the mean PSF used. The scales
and  individual error bars can be compared directly from Figure 2.

2. Azimuthally averaged profiles of QSOs and PSFs from program images.
PSFs are matched at the profile peak, and have error bars less than the
plot symbols in the ranges shown. Magnitude scales are referred to the
limiting magnitude values in Table 1.

3. Contours of R image of 10214+4724. Levels are a factor $\sim$1.5 apart.
Note that the arc extends to near the nucleus, and the inner extension
perpendicular to the arc.

\vfil\eject
\baselineskip 15pt
\centerline{Table 1. Measured and derived quantities}
\vskip 10pt
\settabs 10\columns
\hrule
\vskip 2pt
\hrule
\vskip 3pt
\+QSO &&FWHM &m$_R$ &limit &N/F &m$_{gal}$ &M$_{nuc}^1$ &M$_{gal}^{1,2}$\cr
\+ &&(") &(mag) &(m/"$^2$) &($\pm$factor) \cr
\vskip 3pt
\hrule
\vskip 5pt
\+0820+296$^R$ &&0.6 &18.5 &26.8 &23 (1.5)&~~21.9 &-26.9 &-23.5\cr
\+10214+4724$^R$ &&0.7 &20.2 &25.2 &1.9 (1.5) &~~21.5 &-24.7 &-23.9,-23.9\cr
\+1232+134 &&0.9 &17.4 &26.7 &150 (2.0) &~~22.8 &-28.0 &-22.6\cr
\+1246+005 &&0.7 &17.7 &25.8 &($>$50) &$>$21.9 &-27.7 &($>$-23.5)\cr
\+1338+277 &&0.8 &20.7 &27.7 &4 ~(2.0) &~~22.4 &-24.5 &-23.0\cr
\+1632+391$^{3,R}$ &&0.7 &17.7 &25.5 &35 (1.5) &~~21.6 &-25.4 &-21.5\cr
\+1641+395 &&0.9 &18.6 &26.5 &50 (2.5) &~~23.6 &-26.8 &-21.8,-22.3\cr
\+1641+410 &&0.9 &20.1 &26.1 &15 (2.5) &~~23.6 &-25.3 &-21.8,-21.8\cr
\vskip 5pt
\hrule
\vskip 2pt
\hrule
\vskip 8pt
\centerline{Notes to Table 1}
\vskip 5pt
$^1$ No k-correction. For B0 star k = -2.0m; for A0 star k = 0.5m.

$^2$ Estimates for companion galaxies where separated: not included in N/F
value. Uncertainties are tied to the N/F uncertainty factors quoted.

$^3$ z = 1.1; all others z=2.3.

R = radio-loud source
\end